\documentclass[lettersize,journal]{IEEEtran}
\usepackage{amsmath,amsfonts}
\usepackage{algorithmic}
\usepackage{algorithm}
\usepackage{array}
\usepackage[caption=false,font=normalsize,labelfont=sf,textfont=sf]{subfig}
\usepackage{comment}
\usepackage{textcomp}
\usepackage{stfloats}
\usepackage{url}
\usepackage{verbatim}
\usepackage{graphicx}
\usepackage{cite}
\usepackage{makecell}
\usepackage[normalem]{ulem}
\usepackage{xcolor}

\begin{document}

\title{Realistic modeling of Photonic Terahertz Communication based on a real-world 30 km 30 Gbps experimental demonstration}

\author{Mingxu Wang, Jianjun Yu,~\IEEEmembership{Fellow,~IEEE}, Xianming Zhao, Jiali Chen, Ye Zhou, Xin Lu, Hansong Ma, Chengzhen Bian, Wen Zhou, Kaihui Wang, Weiping Li and Iman Tavakkolnia,~\IEEEmembership{Senior Member,~IEEE}
\thanks{This work was supported in part by the National Key R$\&$D Program of China (Grant No.2023YFB2905600), National Natural Science Foundation of China (No.62127802, 62331004, 62305067, U24B20142, U24B20168, 62427815) and Key Project of Jiangsu Province of China (No.BE2023001-4). (Corresponding author: Jianjun Yu, Wen Zhou and Iman Tavakkolnia).}
\thanks{M. Wang and I. Tavakkolnia are with the Electrical Division, Department of Engineering, University of Cambridge, CB3 0FA, Cambridge, U.K (e-mail:mxwang21@m.fudan.edu.cn; it360@cam.ac.uk)}
\thanks{M. Wang, J. Yu, J. Chen, Y. Zhou, X. Lu, H. Ma, C. Bian, W. Zhou, K. Wang, W. Li are with the State Key Laboratory of ASIC and System, Key Laboratory for Information Science of Electromagnetic Waves (MoE), and the School of Information Science and Technology, Fudan University, Shanghai, 200433, China (e-mail: $\{$mxwang21, jianjun, 23210720141, 24110720111, 23210720218, 23110720118, 24110720124, zwen, khwang, liwp$\}$@m.fudan.edu.cn)}
\thanks{X. Zhao is the Chief Scientist of Beijing Hongshan Information Technology Research Institute Co.,Ltd. Beijing, 100176, China (e-mail: bwcx1998@vip.163.com)}}

\markboth{Journal of \LaTeX\ Class Files,~Vol.~14, No.~8, August~2021}%
{Shell \MakeLowercase{\textit{et al.}}: A Sample Article Using IEEEtran.cls for IEEE Journals}

\IEEEpubid{ }

\maketitle

\begin{abstract}
Experimental studies have demonstrated that THz links can support multi-kilometer, gigabit-per-second wireless transmission. However, few studies have effectively validated channel models through high-capacity, long-distance experiments. In this work, we experimentally demonstrate a 30 km ultra-long-haul photonic THz wireless communication system. 16 Gbaud quadrature phase shift keying (QPSK) signals are transmitted and tested, achieving a maximum data rate of 30 Gbps. For the first time, we validate two widely used THz channel models using a fully implemented 30 km experimental setup, providing a systematic and thorough analysis that integrates theoretical modeling with experimental results. We also perform an analysis of power consumption in photonic THz communication systems and provide a detailed comparison between simulated and experimental power consumption results. Finally, a detailed performance comparison between the simulation and experimental results, including signal-to-noise ratio (SNR), bit error rate (BER) and data rate, are provided. Our work provides essential insights for the design and energy-efficient deployment of future high-capacity THz networks, establishing a benchmark for subsequent research in channel modeling, experimental verification, and system-level energy efficiency.
\end{abstract}

\begin{IEEEkeywords}
THz channel model, photonic THz communication system, model verification, power consumption, consumption factor.
\end{IEEEkeywords}

\section{Introduction}
\IEEEPARstart{W}{ith} the rapid development of technologies such as artificial intelligence (AI), the Internet of Things (IoT), and multimedia services \cite{Ref.1}, future communication infrastructures are expected to support unprecedented traffic volumes, massive connectivity, and stringent latency requirements. As the underlying transport layer of wireless access networks, X-haul, which encompasses fronthaul, midhaul and backhaul, plays a critical role in enabling efficient data aggregation, synchronization, and resource coordination across distributed network nodes. While fifth-generation (5G) systems have established a basic Xhaul framework to support enhanced mobile broadband and centralized radio access network architectures \cite{Ref.3,Ref.5}, the evolution toward sixth-generation (6G) networks is expected to impose far more demanding requirements on Xhaul links, including ultra-high capacity, expanded spectrum utilization, ultra-low latency, high reliability, and flexible scalability to support ultra-dense and heterogeneous network deployments \cite{Ref.6,Ref.7,Ref.9,Ref.10}. Terahertz (THz) technology, operating in the frequency range from 0.1 THz to 1 THz, is one of the key enablers towards this aim. The THz band offers massive bandwidth resources, capable of supporting data rates from tens to thousands of gigabits per second \cite{Ref.12,Ref.13,Ref.14,Ref.15,Ref.16}. In addition, integrating THz wireless links with high-capacity fiber networks \cite{Ref.17,Ref.19,Ref.21,Ref.22} combines the flexibility of wireless communication with the reliability of optical fiber, enabling seamless high-speed transmission across diverse X-haul scenarios.

Channel modeling is the foundation for system design, optimization, and performance analysis. An in-depth investigation of channel modeling in the THz band was presented in \cite{Ref.23}. A comprehensive review of THz-band propagation characteristics, together with sub-THz propagation measurement studies conducted in indoor industrial environments, was reported in\cite{Ref.24}. A simplified path-loss model for the 275–400 GHz frequency range was proposed in \cite{Ref.25} and subsequently adopted in \cite{Ref.26} to assess THz link performance, focusing on metrics such as the average signal-to-noise ratio (SNR) and channel capacity. In \cite{Ref.27}, the authors developed a path-loss formulation that accounts for total absorption loss by considering air, natural gas, and/or water as constituents of the propagation medium, while a multi-ray propagation framework for THz communications was introduced in \cite{Ref.28}. Moreover, a novel THz-band propagation model for electromagnetic nanoscale communications was established in \cite{Ref.29} based on radiative transfer theory, and later extended in \cite{Ref.30} to include the effects of molecular relaxation. Additionally, in \cite{Ref.32,Ref.34}, authors presented suitable stochastic models that are able to accommodate the multipath fading effect in the THz band.

Experimental demonstrations \cite{Ref.37,Ref.38,Ref.40,Ref.41,Ref.42,Ref.44} have shown that THz systems are capable of supporting multi-kilometer, gigabit-per-second wireless transmission. In particular, wireless transceivers operating in the 120 GHz band were developed in \cite{Ref.46}, enabling the successful transmission of 10.3 Gbps amplitude shift keying (ASK) signals over a distance of 5.8 km. A dual-polarization 2 × 2 MIMO transmission system was realized at 125 GHz over a 4.6 km link, delivering a maximum bit rate of 200 Gbps and reaching a record-breaking transmission-distance product of 920 Gbps·km \cite{Ref.47}. Ultra-long-range THz wireless links were experimentally demonstrated in 2017, achieving transmission distances of 21 km and 27 km with data rates of 5 Gbps and 0.5 Gbps, respectively \cite{Ref.48,Ref.49}. Moreover, a 30 km wireless transmission of 4 Gbps QPSK signals at 125 GHz was reported in \cite{Ref.50}, followed by the proposal and experimental validation of a channel response–based maximum ratio combination (CR-MRC) algorithm for orthogonal frequency division multiplexing (OFDM) channel estimation in a 30.4 km THz wireless link, achieving a peak data rate of 20 Gbps \cite{Ref.51}.

Although many studies have been conducted on THz channel modeling and communication experiments, few works have deeply integrated these two aspects by validating and analyzing channel models through high-capacity, long-distance THz communication experiments. Additionally, analyzing the power consumption of THz communication systems is crucial for enabling energy-efficient, scalable, and practically deployable future networks. However, there is currently no comprehensive analysis of the energy consumption of photonic THz communication systems.

In this paper, a 30 km ultra-long-haul high-capacity THz wireless communication system, with a maximum date rate of 30 Gbps, was experimentally demonstrated. For the first time, we compare widely used THz channel models using a fully implemented THz communication setup, providing a systematic analysis of their accuracy. Furthermore, we carry out, also for the first time, an in-depth assessment of power consumption and consumption factor (CF) in photonic THz communication systems and further compare them between simulation and experimental results. These comprehensive analyses offer critical insights for the design and optimization of future high-capacity terahertz networks, establishing a benchmark for subsequent studies in channel modeling, experimental validation, and system-level energy efficiency.

The rest of this paper is organized as follows. Section II illustrates one THz channel model, including path loss, misalignment effects and multipath fading. Section III introduces the transceiver models and power consumption model for photonic THz point-to-point communication system.  Section IV presents the experimental scenarios for the 30 km outdoor THz point-to-point communication system and then compares them with the link model, analyzing the power budget and pointing errors. Section V compares the simulation and experimental results, including SNR, bit error rate (BER), data rate, and other performance metrics. Section VI concludes the paper. Appendix introduces an alternative THz link model and compares it with experimental results.

\section{Channel Model For Outdoor THz Communication System}
\label{sec:II}
In the THz link, the effective channel gain is determined by three dominant degradation factors, as shown in \eqref{eq1}, including path loss $g_\mathrm{a}$, misalignment effects $g_\mathrm{m}$ due to pointing errors, and multipath fading $g_\mathrm{f}$. Hence, the overall channel gain is represented as
\begin{equation}
\label{eq1}
{g} = {g_\mathrm{a}}   {g_\mathrm{m}}   {g_\mathrm{f}}.
\end{equation}

\subsection{Path loss}
The path loss in a THz link can be expressed as
\begin{equation}
\label{eq2}
\begin{alignedat}{1}
g_\mathrm{a} &= g_\mathrm{a\_f} \cdot g_\mathrm{a\_m} \\[2pt]
    &= \frac{c}{4 \pi f d} 
       \, \exp\Biggl[- \frac{1}{2} \kappa_\alpha(f) \, d \Biggr],
\end{alignedat}
\end{equation}
where $g_\mathrm{a\_f}$ denotes free-space path loss (FSPL), $g_\mathrm{a\_m}$ represents the molecular absorption loss \cite{Ref.52}, $c$ represents the speed of light, $f$ is the operating frequency, and $d$ denotes the transmission distance of the THz link. $\kappa_\alpha$($f$) denotes the absorption coefficient describing the relative area per unit of volume, in which the molecules of the medium are capable of absorbing the electromagnetic wave energy. 

\subsection{Misalignment effect}
\begin{figure*}[!t]
\centering
\includegraphics[width=1\linewidth]{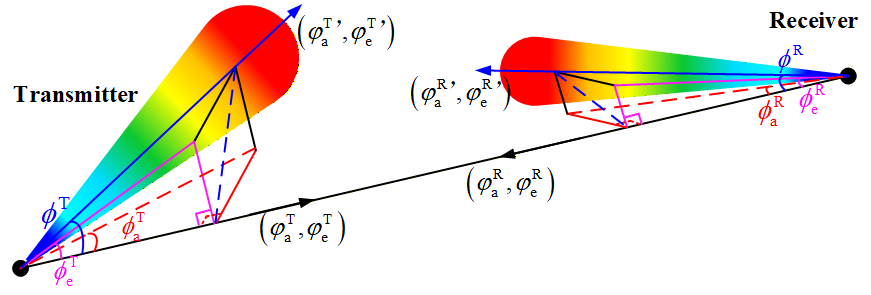}
\caption{Schematic illustration of angle deviation.}
\label{fig1}
\end{figure*}

The misalignment effect in a THz link can be quantified by the pointing errors caused by antenna orientation \cite{Ref.53}, which is expressed as 
\begin{equation}
\label{eq3}
\begin{array}{l}
g_\mathrm{m}
= \sqrt{G_\mathrm{Tmax}G_\mathrm{Rmax}} \,
  \exp\!\left[
    - \frac{1}{2}
    \left( \frac{\phi^\mathrm{T}}{\sigma^\mathrm{T}} \right)^2
    - \frac{1}{2}
    \left( \frac{\phi^\mathrm{R}}{\sigma^\mathrm{R}} \right)^2
  \right] \\[3ex]

= \sqrt{G_\mathrm{Tmax}} \,
  \exp\!\left[
    - \frac{1}{2}
    \left(
      \frac{
        \arctan\!\Bigl(
          \sqrt{ \tan^2(\phi_\mathrm{a}^\mathrm{T}) + \tan^2(\phi_\mathrm{e}^\mathrm{T}) }
        \Bigr)
      }{\Bigl(\sigma^\mathrm{T}\Bigr)}
    \right)^2
  \right] \\[4ex]

\cdot\sqrt{G_\mathrm{Rmax}} \,
\exp\!\left[
  - \frac{1}{2}
  \left(
    \frac{
      \arctan\!\Bigl(
        \sqrt{ \tan^2(\phi_\mathrm{a}^\mathrm{R}) + \tan^2(\phi_\mathrm{e}^\mathrm{R}) }
      \Bigr)
    }{\Bigl(\sigma^\mathrm{R}\Bigr)}
  \right)^2
\right],
\end{array}
\end{equation}
where $\phi_\mathrm{a}^\mathrm{T/R}$ and $\phi_\mathrm{e}^\mathrm{T/R}$ represent the azimuth and elevation angle deviations, as shown in Fig. \ref{fig1}. $\sigma^\mathrm{T/R}$ denotes the half-power beamwidth (HPBW) of the antenna. $G_\mathrm{Tmax}$ and $G_\mathrm{Rmax}$ represent the maximum value of the transmitter and receiver antenna gain, respectively. For small values of $\phi_\mathrm{a}^\mathrm{T/R}$ and $\phi_\mathrm{e}^\mathrm{T/R}$, 
\begin{equation}
\label{eq4}
\begin{aligned}
\phi^\mathrm{T/R}
&= \arctan \!\left(
     \sqrt{
       \tan^2\!\left( \phi_\mathrm{a}^\mathrm{T/R} \right)
       + \tan^2\!\left( \phi_\mathrm{e}^\mathrm{T/R} \right)
     }
   \right) \\[4pt]
&\simeq
   \sqrt{
     \left( \phi_\mathrm{a}^\mathrm{T/R} \right)^2
     + \left( \phi_\mathrm{e}^\mathrm{T/R} \right)^2
   } .
\end{aligned}
\end{equation}
Therefore, \eqref{eq3} can be approximated as
\begin{equation}
\label{eq5}
\begin{aligned}
g_\mathrm{m}
&\simeq
\sqrt{
  G_\mathrm{Tmax} \,
  \exp\!\left(
    - \frac{
      (\phi_\mathrm{a}^\mathrm{T})^2 + (\phi_\mathrm{e}^\mathrm{T})^2
    }{
      (\sigma^\mathrm{T})^2
    }
  \right)
} \\[4pt]
&\quad \cdot
\sqrt{
  G_\mathrm{Rmax} \,
  \exp\!\left(
    - \frac{
      (\phi_\mathrm{a}^\mathrm{R})^2 + (\phi_\mathrm{e}^\mathrm{R})^2
    }{
      (\sigma^\mathrm{R})^2
    }
  \right)
}.
\end{aligned}
\end{equation}

As demonstrated in Fig. \ref{fig1}, we define $\varphi_\mathrm{a}^\mathrm{T/R}$ and $\varphi_\mathrm{e}^\mathrm{T/R}$ as the azimuth and elevation angles from the center of the transmitter antenna to the center of the receiver antenna, respectively, which can be regarded as the ideal pointing angles. The actual azimuth and elevation angles of the antenna are denoted by ${\varphi_\mathrm{a}^\mathrm{T/R}}’$ and ${\varphi_\mathrm{e}^\mathrm{T/R}}’$ respectively, as shown in Fig. \ref{fig1}. Note that $\varphi$ represents the absolute angle, whereas $\phi$ denotes the relative angle, i.e., the angle deviation. Additionally, $\sigma_\mathrm{a}^\mathrm{T/R}$ and $\sigma_\mathrm{e}^\mathrm{T/R}$ denote the HPBW of the antenna in the azimuth and elevation directions, respectively. Therefore, misalignment effect can be finally expressed as \cite{Ref.54,Ref.55}

\begin{equation}
\label{eq6}
\begin{array}{l}
g_\mathrm{m} \simeq \sqrt{G_\mathrm{Tmax}}
\sqrt{\exp\!\Biggl[
- \left( \frac{\varphi_\mathrm{a}^\mathrm{T} - \varphi_\mathrm{a}^\mathrm{T'}}{\sigma_\mathrm{a}^\mathrm{T}} \right)^2
- \left( \frac{\varphi_\mathrm{e}^\mathrm{T} - \varphi_\mathrm{e}^\mathrm{T'}}{\sigma_\mathrm{e}^\mathrm{T}} \right)^2
\Biggr]} \\[1.5ex]
\cdot\sqrt{G_\mathrm{Rmax} }
\sqrt{\exp\!\Biggl[
- \left( \frac{\varphi_\mathrm{a}^\mathrm{R} - \varphi_\mathrm{a}^\mathrm{R'}}{\sigma_\mathrm{a}^\mathrm{R}} \right)^2
- \left( \frac{\varphi_\mathrm{e}^\mathrm{R} - \varphi_\mathrm{e}^\mathrm{R'}}{\sigma_\mathrm{e}^\mathrm{R}} \right)^2
\Biggr]} .
\end{array}
\end{equation}

\subsection{Multipath fading}
\label{sec:II.C}
The multipath fading effect $g_\mathrm{f}$ of THz link is modeled as a generalized $\alpha-\mu$ distribution \cite{Ref.56} and its probability density function (PDF) can be expressed as 
\begin{equation}
\label{eq7}
f_{g_\mathrm{f}}(g_\mathrm{f}) =
\frac{\alpha \, \mu^\mu}{\widehat{g_\mathrm{f}}^{\alpha \mu} \, \Gamma(\mu)} \,
g_\mathrm{f}^{\alpha \mu - 1} \,
\exp\Biggl[
- \mu \, \frac{g_\mathrm{f}^\alpha}{\widehat{g_\mathrm{f}}^\alpha}
\Biggr],
\end{equation}
where $\alpha >$ 0 denotes the fading parameter, $\mu$ represents the normalized variance of the channel envelope, and $\widehat{{g_\mathrm{f}}}$ is the $\alpha$-root mean value of the fading envelope. It is worth noting that this distribution serves as a generalized model encompassing several classical fading distributions, including Rayleigh ($\alpha$ = 2, $\mu$ = 1), Nakagami-m ($\alpha$ = 2 and $\mu$ is the fading parameter), and Weibull ($\mu$ = 1 with $\alpha$ as the fading parameter), among others.

Note that there is another widely used THz link model which will be presented and discussed in the Appendix. We find that the model in the Appendix is usually used wrongly in the literature. 

\section{Transceiver Model For Photonic THz Communication System}
\begin{figure*}[!t]
\centering
\includegraphics[width=1\linewidth]{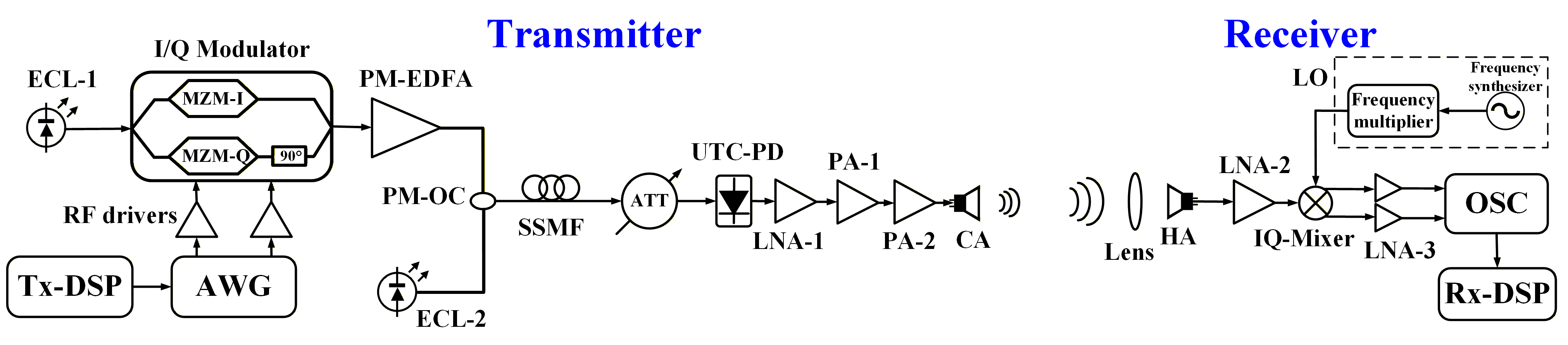}
\caption{Schematics of photonic THz transceiver. AWG: arbitrary waveform generator, RF drivers: radio frequency drivers, ECL: external cavity laser, I/Q Modulator: in-phase/quadrature modulator, PM-EDFA: polarization-maintaining erbium-doped fiber amplifier, PM-OC: polarization-maintaining optical coupler, SSMF: standard single-mode fiber, ATT: attenuator, UTC-PD: unitraveling-carrier photodiode, LNA: low-noise amplifier, PA: power amplifier, CA: Cassegrain antenna, HA: horn antenna, LO: local oscillator, OSC: oscilloscope, DSP: digital signal processing.}
\label{fig2}
\end{figure*}

The schematics of photonic THz transceiver is shown in Fig. 2. The full names of the components are provided in the caption of Fig. 2. Since the setup in this paper is similar to that in Ref. [38], it is not described in detail here; only the differences between the two are elaborated. First, the PA-1 model and gain are different. In the setup used in Ref. \cite{Ref.51}, PA-1 (AT-PA-110150-3013) has a gain of 30 dB, which drives the cascaded PA-2 into deep saturation. Although this increases the transmitting power, it also leads to more nonlinear effects. In the current setup, PA-1 (AT-PA-85140-1310) has a gain of 13 dB, which mitigates nonlinear effects and improves linearity. Second, in Ref. \cite{Ref.51}, the receiver uses a balanced mixer (SFB-06-N1) and down-converted the THz signal to intermediate frequency (IF) signal, whereas the current setup employs an IQ mixer (TMIQ-120160-0230-06) and directly generates a complex baseband signal. This effectively avoids image interference before OSC, reduces bandwidth and dynamic range requirements, and simplifies.

\subsection{SNR of THz system}
The noise variance \cite{Ref.58} of photonic THz communication system is given by
\begin{equation}
\label{eq8}
\sigma _\mathrm{THz}^2 = k   \left( {{T_\mathrm{ant}} + \left( {N{F_\mathrm{R}} - 1} \right)   {T_0}} \right)   B,
\end{equation}
where $k$ represents the Boltzmann constant (1.38$\times$10$^{-23}$ J/K), $T_{0}$ denotes the reference temperature (290 K), $T_\mathrm{ant}$ is antenna equivalent noise temperature, which represents the thermal noise contributed by the atmosphere and background radiation, $NF_\mathrm{R}$ is the noise figure of the THz receiver expressed in the linear domain, $B$ is the signal bandwidth.

The Friis formula for noise factor is used to calculate $NF_\mathrm{R}$ of THz receiver as follow
\begin{equation}
\label{eq9}
N{F_\mathrm{R}} = N{F_\mathrm{LNA - 2}} + \frac{{N{F_\mathrm{Mixer}} - 1}}{{{G_\mathrm{LNA - 2}}}} + \frac{{N{F_\mathrm{LNA - 3}} - 1}}{{{G_\mathrm{LNA - 2}}   {G_\mathrm{Mixer}}}},
\end{equation}
where $NF_\mathrm{LNA-2}$ and $G_\mathrm{LNA-2}$ denote the noise figure and gain of LNA-2, respectively, $NF_\mathrm{Mixer}$ and $G_\mathrm{Mixer}$ denote the noise figure and gain of mixer, which can be approximated as the conversion loss of the mixer, and $NF_\mathrm{LNA-3}$ denotes the noise figure of LNA-3. Note that all the parameters in \eqref{eq9} are expressed in the linear domain. Assuming $T_\mathrm{ant}$ = $T_{0}$, the SNR of the photonic THz communication system is expressed as
\begin{equation}
\label{eq10}
{\gamma} = \frac{{{P_\mathrm{R}}}}{{k   {T_0}   (N{F_\mathrm{R}})   B}}.
\end{equation}
where $P_\mathrm{R}$ denotes the received power. 

\subsection{Consumption factor of photonic THz system}
The CF, which is a well-known parameter used in the literatures, is defined as the ratio of data rate to power consumed\cite{Ref.59},
\begin{equation}
\label{eq11}
C{F} = \frac{{R}}{{P_\mathrm{consumed}}},
\end{equation}
where $R$ represents the data rate and $P_\mathrm{consumed}$ represents the total power consumption of photonic THz communication system. The unit of CF is bps/W. CF can also be defined as number of bits that can be transmitted through a communication system for every Joule of expended energy, and it unit is bits/J. In the simulations, the achievable data rate is computed using the Shannon capacity in order to evaluate the theoretical upper bound of system performance, which can be expressed as
\begin{equation}
\label{Shannon}
R = B{{\log }_2}\left( {1 + {\gamma}} \right).
\end{equation}
In the experiment, however, the data rate is determined by the actual transmission parameters, i.e., the occupied bandwidth multiplied by the spectral efficiency $\eta$ (bps/Hz) of the implemented modulation, which can be expressed as
\begin{equation}
\label{rate_SE}
R = \eta B.
\end{equation}
Although the rate is obtained using different expressions, both approaches quantify the effective information throughput under the corresponding conditions, and thus the CF metric remains physically consistent. Additionally, $P_\mathrm{consumed}$ consists of $P_\mathrm{Tx\_consumed}$ and $P_\mathrm{Rx\_consumed}$, and they are expressed as \cite{Ref.59}
\begin{equation}
\label{eq12}
\begin{cases}
\begin{aligned}
P_\mathrm{Tx\_consumed} &= \frac{P_\mathrm{T}}{h_\mathrm{cascaded}} + P_\mathrm{AWG} + 2 \, P_\mathrm{RF} + 2 \, P_\mathrm{ECL} \\
&\quad + P_\mathrm{Mod} + P_\mathrm{EDFA} + P_\mathrm{PD},\\[1.5ex]
P_\mathrm{Rx\_consumed} &= P_\mathrm{LNA-2} + P_\mathrm{Mixer} + P_\mathrm{Syn} + P_\mathrm{Mul} + \\
&\quad 2 \, P_\mathrm{LNA-3} + P_\mathrm{OSC}.
\end{aligned}
\end{cases}
\end{equation}
where $P_\mathrm{AWG}$, $P_\mathrm{RF}$, $P_\mathrm{ECL}$, $P_\mathrm{Mod}$, $P_\mathrm{EDFA}$, $P_\mathrm{PD}$, $P_\mathrm{LNA-2}$, $P_\mathrm{LNA-3}$, $P_\mathrm{Mixer}$, $P_\mathrm{Syn}$, $P_\mathrm{Mul}$, $P_\mathrm{OSC}$ denote the power consumed by AWG, RF driver, ECL, I/Q modulator, PM-EDFA, UTC-PD, LNA-2, LNA-3, IQ-subharmonic mixer, frequency synthesizer, frequency multiplier and OSC. $P_\mathrm{T}$ represents the transmitted power, $h_\mathrm{cascaded}$ denotes the power-efficiency factor of cascaded electronic components in the transmitter, including LNA-1, PA-1 and PA-2. It is defined as \cite{Ref.59}
\begin{equation}
\label{eq13}
{h_\mathrm{cascaded}} = {\left[ {1 + \sum\limits_{p = 1}^M {\frac{1}{{\mathop \Pi \limits_{q = p + 1}^M {G_q}}}   \left( {\frac{1}{{{\eta _p}}} - 1} \right)} } \right]^{ - 1}},
\end{equation}
where $G_{q}$ is the gain of the $q-\mathrm{th}$ component, $M$ is the number of cascaded components and $\eta_{p}$ is the efficiency of the $p-\mathrm{th}$ component, which is defined as 
\begin{equation}
\label{eq14}
{\eta _p} = \frac{{{P_{\mathrm{sig}\_p}}}}{{{P_{\mathrm{sig}\_p}} + {P_{\mathrm{consumed}\_p}}}},
\end{equation}
where $P_{\mathrm{sig}\_p}$ is the signal power delivered from the $p-\mathrm{th}$ component to $(p+1)-\mathrm{th}$ component, and the $P_{\mathrm{consumed}\_p}$ denotes the non-signal power consumed by the $p-\mathrm{th}$ component. For the photonic THz transmitter in Fig. \ref{fig2}, $P_\mathrm{T}$ is equal to the output power of PA-2, and $h_\mathrm{cascaded}$ is expressed as 
\begin{equation}
\label{eq15}
h_\mathrm{cascaded} =
\Biggl[
\begin{aligned}
& 1 
+ \frac{1}{G_\mathrm{PA-1} G_\mathrm{PA-2}} \left( \frac{1}{\eta_\mathrm{LNA-1}} - 1 \right)+  \\[1.2ex]
& \frac{1}{G_\mathrm{PA-2}} \left( \frac{1}{\eta_\mathrm{PA-1}} - 1 \right) 
+ \left( \frac{1}{\eta_\mathrm{PA-2}} - 1 \right)
\end{aligned}
\Biggr]^{-1}
\end{equation}
where $\eta_\mathrm{LNA-1}$, $\eta_\mathrm{PA-1}$ and $\eta_\mathrm{PA-2}$ are expressed as
\begin{equation}
\label{eq16}
\begin{cases}
\eta_\mathrm{LNA-1} = \dfrac{P_\mathrm{LNA-1\_out}}{P_\mathrm{LNA-1\_out} + P_\mathrm{LNA-1}},\\[3ex]
\eta_\mathrm{PA-1}  = \dfrac{P_\mathrm{PA-1\_out}}{P_\mathrm{PA-1\_out} + P_\mathrm{PA-1}},\\[3ex]
\eta_\mathrm{PA-2}  = \dfrac{P_\mathrm{T}}{P_\mathrm{T} + P_\mathrm{PA-2}}.
\end{cases}
\end{equation}
where $P_\mathrm{LNA-1}$, $P_\mathrm{PA-1}$, and $P_\mathrm{PA-2}$ denotes the power consumed by LNA-1, PA-1 and PA-2, respectively, $P_\mathrm{LNA-1\_out}$ represents the output power of LNA-1, $P_\mathrm{PA-1\_out}$ represents the output power of PA-1. Note that PM-OC, ATT, CA, lens and HA are passive components. Without loss of generality, we do not include the power consumption of the offline DSP, since it is performed on a laptop. This does not affect the final results, because this part is not considered in either the simulation or the experiment. If the DSP were implemented on the field-programmable gate array (FPGA), its power consumption would be easier to quantify, which will be an aspect to be addressed in future work.

In an AWG, only the power consumed by the digital-to-analog converter (DAC) is considered. According to \cite{Ref.60}, the power consumption of the DAC is approximated by
\begin{equation}
\label{eq17}
{P_\mathrm{AWG}} = 2 \cdot{F_\mathrm{DAC}}\cdot {r_\mathrm{DAC}}  \cdot {F_\mathrm{S\_DAC}},
\end{equation}
where $F_\mathrm{DAC}$ is a DAC figure of merit, $r_\mathrm{DAC}$ is the DAC resolution, $F_\mathrm{S\_DAC}$ is the sampling frequency and ‘2’ denotes two DAC channels due to I/Q modulation.

The power consumed by the high-speed I/Q modulator, which is typically traveling-wave devices with internal resistive terminations, is given by 
\begin{equation}
\label{eq18}
{P_\mathrm{Mod}} = 4   \left( {{V_\mathrm{bias}} \cdot  {V_\mathrm{RF}}} \right)/{R_\mathrm{T}},
\end{equation}
where $V_\mathrm{bias}$ is the bias voltage, $V_\mathrm{RF}$ is the peak-to-peak modulation voltage and $R_\mathrm{T}$ is the termination resistance for the driver as well as the Mach-Zehnder interferometer (MZI) \cite{Ref.60,Ref.61}.

The power consumed by the UTC-PD is given by \cite{Ref.60,Ref.61,Ref.62}
\begin{equation}
\label{eq19}
{P_\mathrm{PD}} =  {R_\mathrm{PD}} \cdot  \left| {V_\mathrm{PD}} \right|  \cdot {P_\mathrm{in}},
\end{equation}
where $R_\mathrm{PD}$ is the responsivity of the UTC-PD, $V_\mathrm{PD}$ is the bias voltage of the UTC-PD and $P_\mathrm{in}$ denotes the input optical power of the UTC-PD.

In an OSC, only the power consumed by the analog-to-digital converter (ADC) is considered. According to \cite{Ref.60}, the power consumption of the ADC is approximated by
\begin{equation}
\label{eq20}
{P_\mathrm{OSC}} = 2  \cdot {F_\mathrm{ADC}} \cdot  {r_\mathrm{ADC}} \cdot  {F_\mathrm{S\_ADC}},
\end{equation}
where $F_\mathrm{ADC}$ is a ADC figure of merit, $r_\mathrm{ADC}$ is the ADC resolution, $F_\mathrm{S\_ADC}$ is the sampling frequency and ‘2’ denotes that two ADC channels are used due to the coherent reception.

\section{Experimental Setup For Outdoor Photonic THz Link}
\label{sec:IV}

The key specifications of the transmitter and receiver are summarized in Table \ref{table1}.

\begin{table*}[!t]
\caption{Key Specifications of the Transceiver\label{table1}}
\centering
\renewcommand{\arraystretch}{3.2}
\begin{tabular}{cccc}
\hline
\textbf{Component} & \textbf{Specification} & \textbf{\makecell*[c]{Power consumption\\(simulation)}} & \textbf{\makecell*[c]{Power consumption\\(experiment)}}\\
\hline
\makecell*[c]{AWG\\(Keysight M8195A)} & \makecell*[c]{Sampling rate ($F_\mathrm{S\_DAC}$): 65 GSa/s\\Resolution ($r_\mathrm{DAC}$): 8 bit\\Figure of merit ($F_\mathrm{DAC}$): 1.7$\times$10$^{-12}$J/conv-step} & \makecell*[c]{1.77 W\\ (2$\times$65$\times$10$^{9}$$\times$8$\times$1.7$\times$10$^{-12}$)}& \makecell*[c]{1.77 W$^{*}$\\(2$\times$65$\times$10$^{9}$$\times$8$\times$1.7$\times$10$^{-12}$)}\\
\hline
Electrical Attenuator & Gain: -14 dB & 0 W & 0 W\\
\hline
\makecell*[c]{RF driver\\(SHF S804 B)} & Gain: 22 dB & \makecell*[c]{3.96 W \\(2$\times$9 V$\times$0.22 A)} & \makecell*[c]{3.78 W \\(2$\times$8.6 V$\times$0.22 A)}\\
\hline
\makecell*[c]{ECL-1/2\\(ITU-M-C-15-P-FA)} & \makecell*[c]{Wavelength-1/2: 1550/1551.024 nm\\Linewidth: 100 kHz} & \makecell*[c]{6 W \\(2$\times$5 V$\times$0.6 A)} & \makecell*[c]{6 W$^{*}$ \\(2$\times$5 V$\times$0.6 A)}\\
\hline
\makecell*[c]{IQ modulator\\(FTM7961EX)} & \makecell*[c]{Modulation voltage ($V_\mathrm{RF}$): 2.52 V\\Half-wave voltage ($V_\pi$): 3.5 V\\Bias voltage ($V_\mathrm{bias}$): 3.5 V\\RF impedance ($R_\mathrm{T}$): 50 $\Omega$} & \makecell*[c]{0.706 W \\(4$\times$2.52 V$\times$3.5 V/50 $\Omega$)} & \makecell*[c]{0.05 W \\(2$\times$10.6 V$\times$0.002 A+3.8 V$\times$0.002 A)}\\
\hline
\makecell*[c]{PM-EDFA\\(EDFA-D-C-20-0-P-FA)} & \makecell*[c]{Gain: 23 dB \\Noise figure: 5.5 dB} & \makecell*[c]{5 W \\(5 V$\times$1 A)} & \makecell*[c]{5 W$^{*}$ \\(5 V$\times$1 A)}\\
\hline
\makecell*[c]{UTC-PD\\(IOD-PMD-14001)} & \makecell*[c]{Responsivity ($R_\mathrm{PD}$): 0.41 A/W \\Bias voltage ($V_\mathrm{PD}$): -1 V\\Input power: 0 dBm} & \makecell*[c]{4.1$\times$10$^{-4}$ W \\(0.41 A/W$\times$1$\times$10$^{-3}$ W$\times$1 V)}& \makecell*[c]{4.0$\times$10$^{-4}$ W \\(0.41 A/W$\times$1$\times$10$^{-3}$ W$\times$0.98 V)}\\
\hline
\makecell*[c]{LNA-1\\(AT-LNA-110170-1806T)} & \makecell*[c]{Frequency: 110$\sim$170 GHz \\Gain ($G_\mathrm{LNA-1}$): 15 dB} & \makecell*[c]{0.15 W \\(5 V$\times$0.03 A)} & \makecell*[c]{0.117 W \\(4.5 V$\times$0.026 A)}\\
\hline
\makecell*[c]{PA-1\\(AT-PA-85140-1310)} & \makecell*[c]{Frequency: 85$\sim$140 GHz \\Gain ($G_\mathrm{PA-1}$): 13 dB} & \makecell*[c]{0.9 W \\(5 V$\times$0.18 A)} & \makecell*[c]{0.81 W \\(4.5 V$\times$0.18 A)}\\
\hline
\makecell*[c]{PA-2\\(AT-PA-85135-1823)} & \makecell*[c]{Frequency: 85$\sim$135 GHz \\Gain ($G_\mathrm{PA-2}$): 20 dB (Typical)\\Saturated output power: 23 dBm} & \makecell*[c]{14.5 W \\(10 V$\times$1.45 A)} & \makecell*[c]{12.52 W \\(8.6 V$\times$1.456 A)}\\
\hline
\makecell*[c]{CA\\(CRAS-D06-D50G54)} & \makecell*[c]{Frequency: 110$\sim$170 GHz \\Gain ($G_\mathrm{T}$): 54 dBi\\Diameter: 50 cm} & 0 W & 0 W\\
\hline
Lens + HA & Gain ($G_\mathrm{R}$): 57 dBi& 0 W & 0 W\\
\hline
\makecell*[c]{LNA-2\\(AT-LNA-110170-3306E)} & \makecell*[c]{Frequency: 110$\sim$170 GHz \\Gain ($G_\mathrm{LNA-2}$): 38 dB\\Noise figure ($NF_\mathrm{LNA-2}$): 3.4 dB} & \makecell*[c]{0.35 W \\(5 V$\times$0.07 A)} & \makecell*[c]{0.34 W \\(4.86 V$\times$0.07 A)}\\
\hline
\makecell*[c]{Frequency synthesizer \\(TLFS00522-200)} & \makecell*[c]{Frequency: 0.05$\sim$22.6 GHz \\Output Power: 4 $\pm$ 5 dBm} & \makecell*[c]{8.4 W \\(12 V$\times$0.7 A)} & \makecell*[c]{7.2 W \\(10.5 V$\times$0.686 A)}\\
\hline
\makecell*[c]{Frequency multiplier \\(TMAM-055090-0612-12)} & \makecell*[c]{Frequency: 9.16$\sim$15 GHz \\Multiply factor: 6\\Output frequency: 55$\sim$90 GHz\\Output Power: 12 dBm} & \makecell*[c]{2.52 W \\(12 V$\times$0.21 A)} & \makecell*[c]{2.05 W \\(10.8 V$\times$0.19 A)}\\
\hline
\makecell*[c]{IQ-subharmonic mixer \\(TMIQ-120160-0230-06)} & \makecell*[c]{RF: 120$\sim$160 GHz\\LO: 60$\sim$80 GHz\\IF: DC$\sim$30 GHz\\Conversion Loss ($G_\mathrm{Mixer}$): -11 dB} & 0 W & 0 W\\
\hline
\makecell*[c]{LNA-3 \\(AT-LNA-0043-3504Y)} & \makecell*[c]{Frequency: 75 kHz$\sim$43.5 GHz\\Gain ($G_\mathrm{LNA-3}$): 35 dB\\Noise figure ($NF_\mathrm{LNA-3}$): 4.5 dB} & \makecell*[c]{3.3 W \\(2$\times$5 V$\times$0.33 A)} & \makecell*[c]{2.88 W \\(2$\times$4.36 V$\times$0.33 A)}\\
\hline
\makecell*[c]{OSC\\(Tektronix DSA 73304D)} & \makecell*[c]{Sampling rate ($F_\mathrm{S\_ADC}$): 100 GSa/s\\Resolution ($r_\mathrm{ADC}$): 8 bit\\Figure of merit ($F_\mathrm{ADC}$): 2.6$\times$10$^{-12}$J/conv-step} & \makecell*[c]{4.16 W\\(2$\times$100$\times$10$^{9}$$\times$8$\times$2.6$\times$10$^{-12}$)} & \makecell*[c]{4.16 W$^{*}$\\(2$\times$100$\times$10$^{9}$$\times$8$\times$2.6$\times$10$^{-12}$)}\\
\hline
\end{tabular}\\
\vspace{2mm}
{\footnotesize $^{*}$ indicates that the power consumption of the component could not be measured experimentally and the value from the datasheet is used instead.}
\end{table*}

\subsection{Experimental scenarios}
\begin{figure*}[!t]
\centering
\includegraphics[width=1\linewidth]{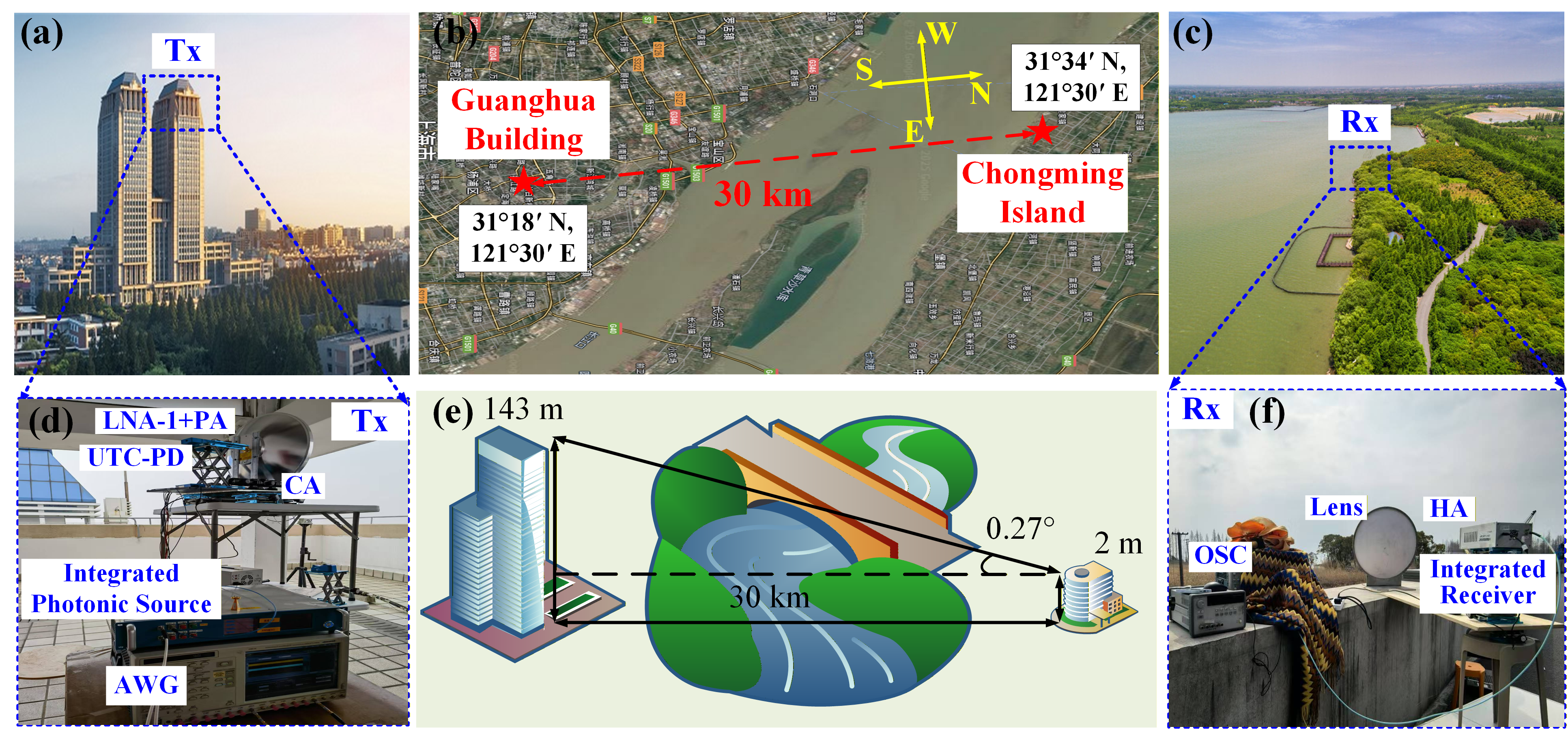}
\caption{Experimental scenarios: (a) photograph of Guanghua Building, (b) locations of the transmitter and receiver, (c) photograph of Chongming Island shoreline, (d) photograph of the transmitter setup, (e) elevation angles of the transmitter and receiver, (f) photograph of the receiver setup.}
\label{fig3}
\end{figure*}

Fig. \ref{fig3}(b) presents experimental scenarios of the 30 km low-altitude THz communication link, comprising an 11 km terrestrial section and a 19 km sea-surface section. The transmitter is mounted on the top floor of the Guanghua Building at Fudan University at an elevation of 143 m, as shown in Fig. \ref{fig3}(a) and (d), while the receiver is deployed on the Chongming Island shoreline in Shanghai city at a height of 2 m, as shown in Fig. \ref{fig3}(c) and (f). This geometry results in a transmitter elevation angle of approximately -0.27° and a receiver elevation angle of approximately 0.27°, as depicted in Fig. \ref{fig3}(e). The azimuth angles of the transmitter and receiver are determined from their latitude and longitude coordinates. As shown in Fig. \ref{fig3}(b), The latitude and longitude coordinates of the transmitter and receiver are (31°18’N, 121°30’E) and (31°34’N, 121°30’E), respectively. Since the transmitter and receiver are very close in longitude, the transmitter can be pointed toward true north, with its azimuth angle defined as 0°, and the receiver can be pointed toward true south, with its azimuth defined as 180°. During the calibration process, the receiver is initially identified through the transmitter’s telescope. The transmitter is carefully pointed toward the receiver, and subsequently, the receiver’s pointing direction and location are adjusted slightly to maximize the received signal power.

\subsection{Signal power analysis of the transmitter}
The AWG generates I- and Q-path signals with a peak-to-peak voltage ($V_\mathrm{pp}$) of 1 V. The modulation format is quadrature phase shift keying (QPSK). After traversing a -14 dB electrical attenuator and a 22 dB RF driver, the signals in both channels exhibit an increase in $V_\mathrm{pp}$ to 2.52 V. These signals are subsequently applied to the I/Q modulator as the driving inputs, and the ECL-1 generates an optical carrier at 1550 nm with an output power of 11 dBm. The output power of the I/Q modulator is -9 dBm. The modulated optical signal is then amplified by a PM-EDFA driven by a pump current of 120 mA, resulting in an output power of 11 dBm. Subsequently, the amplified optical signal is combined with another optical carrier from ECL-2 using a PM-OC. ECL-2 operates at 1551.024 nm and provides an output power of 11 dBm. Note that in this setup, the output powers of PM-EDFA and ECL-2 are adjusted to be equal, thereby maximizing the beat-frequency generation efficiency and THz output power in the UTC-PD. The power of the combined optical signal is 14 dBm. After transmission over a 10 km span of SSMF with 0.2 dB/km attenuation, the optical power decreases to 12 dBm. An optical attenuator is then employed to adjust the incident optical power into the UTC-PD, thereby enabling the generation of THz signals with different output powers. The frequency spacing between the two optical signals is 128 GHz, generating a corresponding 128 GHz THz signal through the UTC-PD.

The derivation of the THz signal power generated by the UTC-PD is as follows. Assuming that the optical carrier power $P_\mathrm{LO}$ and the signal power $P_\mathrm{Sig}$ are both set as $P$/2 (W). The photocurrent generated by the UTC-PD consists of a direct current (DC) component and the desired THz component, expressed as follow
\begin{equation}
\label{eq21}
\begin{array}{l}
{I_\mathrm{PD}}\left( t \right) = {R_\mathrm{PD}}   {\left| {{E_\mathrm{LO}} + {E_\mathrm{Sig}}} \right|^2}\\[1.5ex]
{\rm{          }} = \underbrace {{R_\mathrm{PD}}   \left( {{P_\mathrm{LO}} + {P_\mathrm{Sig}}} \right)}_\mathrm{DC} + \underbrace {2   {R_\mathrm{PD}}   \sqrt {{P_\mathrm{LO}}   {P_\mathrm{Sig}}}    cos\left( {\Delta \omega    t + \Delta \phi } \right)}_\mathrm{Signal}\\[2ex]
{\rm{          }} = {R_\mathrm{PD}}   P + {R_\mathrm{PD}}   P   cos\left( {\Delta \omega    t + \Delta \phi } \right),{\rm{       }}
\end{array}
\end{equation}
where $R_\mathrm{PD}$ is the responsivity, $\Delta\omega$ denotes the frequency of the THz signal and $\Delta\phi$ represents the phase difference. The THz current has a root-mean-square (RMS) value of
\begin{equation}
\label{eq22}
{I_\mathrm{THz,RMS}} = \frac{{{R_\mathrm{PD}}   P}}{{\sqrt 2 }},
\end{equation}
and power of the THz component is
\begin{equation}
\label{eq23}
\left\{
\begin{array}{l}
P_\mathrm{THz}\left( {\rm W} \right) = \left( I_\mathrm{THz,RMS} \right)^2 R_\mathrm{T}
= \dfrac{R_\mathrm{PD}^2 P^2 R_\mathrm{T}}{2}, \\[2mm]
P_\mathrm{THz}\left( {\rm dBm} \right) = 10 \, {\log }_{10} \left( P_\mathrm{THz}\left( {\rm W} \right) \right) + 30\\[2mm]
\quad\quad\quad\quad\quad\;\;  = 10 \, {\log }_{10} \left( \dfrac{R_\mathrm{PD}^2 P^2 R_\mathrm{T}}{2} \right) + 30.
\end{array}
\right.
\end{equation}
where $R_\mathrm{T}$ is the load resistance. 

In the experiment, $R_\mathrm{PD}$ is 0.41 A/W and $R_\mathrm{T}$ is 50 $\Omega$. The incident optical power of the UTC-PD is adjusted using an optical ATT within the range of -5 to 0 dBm, corresponding to 3.16$\times$10$^{-4}\sim 1\times$10$^{-3}$ W. According to \eqref{eq23}, the generated THz signal power ranges from -33.7 to -23.7 dBm. To enable ultra-long-distance wireless transmission, the generated THz signal is amplified by LNA-1 with a gain of 15 dB, boosted by PA-1 with a gain of 13 dB and finally boosted by PA-2, which offers a saturated output power of 23 dBm. The amplifiers’ insertion losses have already been incorporated into gains. Therefore, the transmitting power $P_\mathrm{T}$ is calculated to range from 17.7 dBm (= -33.7+15+13+23.4) to 21.2 dBm (= -23.7+15+13+16.9), as shown in Table \ref{table2}. Note that PA-2 operates in the nonlinear region, and its gain depends on the input signal power. As the input power increases, the output power approaches the saturation level. This explains why the THz signal generated by the UTC-PD exhibits a power variation of 10 dB, whereas the corresponding variation in the final $P_\mathrm{T}$ is limited to only 3.5 dB. Finally, the amplified THz signal is radiated into free space through a CA with a gain of 54 dBi.

\begin{table}[!t]
\caption{Calculation of the transmitting power\label{table2}}
\centering
\renewcommand{\arraystretch}{1.8}
\begin{tabular}{
>{\centering\arraybackslash}m{1.0cm}
>{\centering\arraybackslash}m{2.3cm}
>{\centering\arraybackslash}m{1.9cm}
>{\centering\arraybackslash}m{2.1cm}}
\hline
\textbf{Component} & \textbf{Input power} & \textbf{Gain} & \textbf{Output power}\\
\hline
LNA-1 & -33.7 $\sim$ -23.7 dBm & 15 dB & -18.7 $\sim$ -8.7 dBm \\
\hline
PA-1 & -18.7 $\sim$ -8.7 dBm & 13 dB & -5.7 $\sim$ 4.3 dBm\\
\hline
PA-2 & -5.7 $\sim$ 4.3 dBm & 23.4 $\sim$ 16.7 dB & 17.7 $\sim$ 21.2 dBm\\
\hline
\end{tabular}
\end{table}  

\subsection{Model-based and experiment-based calculation of received power $P_\mathrm{R}$}
\label{sec:IV.C)}
Based on the THz link model presented in Section \ref{sec:II}, the power of the received signal can be estimated as follow,
\begin{equation}
\label{eq24}
\begin{array}{l}
P_\mathrm{R}~(\mathrm{dBm}) = 10 \log_{10} \Bigl[ P_\mathrm{T}~(\mathrm{W}) \cdot (g_\mathrm{a} g_\mathrm{m} g_\mathrm{f})^2 \Bigr] \\[1.5ex]
= P_\mathrm{T} + 20 \log_{10} (g_\mathrm{a}) + 20 \log_{10} (g_\mathrm{m}) + 20 \log_{10} (g_\mathrm{f}) \\[1.5ex]
= P_\mathrm{T} + \underbrace {FSPL+L_\mathrm{m}}_{g_\mathrm{a}} + \underbrace {L_\mathrm{f}}_{g_\mathrm{f}}\\ + \underbrace {G_\mathrm{Tmax} + G_\mathrm{Rmax} + L_\mathrm{p}}_{g_\mathrm{m}} .
\end{array}
\end{equation}
Firstly, in the experiment, $f$ is 128 GHz, $c$ is 3$\times$10$^{8}$ m/s and $d$ is 30 km. Therefore, the FSPL $g_\mathrm{a\_f}$ is 6.22$\times$10$^{-9}$, expressed in the linear domain, corresponding to -164.1 dB. The molecular absorption loss $g_\mathrm{a\_m}$ is calculated based on weather conditions during the experiment. In Shanghai, between 22:30 PM and 23:30 PM on November 21, 2025, the average temperature was 9.8 °C, the average atmospheric pressure was 1028 hPa, the relative humidity was 64$\%$, and the water vapor density was 5.9 g/m$^{3}$. Therefore, $g_\mathrm{a\_m}$ is 0.083 at 128 GHz for the 30 km wireless transmission, corresponding to $L_\mathrm{m}$ = -21.6 dB. Therefore, $g_\mathrm{a}$ is calculated to be 5.14$\times$10$^{-10}$, corresponding to -185.7 dB. Secondly, in the experiment, $G_\mathrm{Tmax}$ and $G_\mathrm{Rmax}$ are 54 dBi and 57 dBi, respectively. Assuming that there is no pointing errors loss $L_\mathrm{p}$ between the transmitter and receiver antenna, $g_\mathrm{m}$ is calculated to be 3.55$\times$10$^{5}$, corresponding to 111 dB. Thirdly, in a point-to-point THz link with highly directional antennas and negligible multipath components, the small-scale fading is almost absent due to the dominance of the line-of-sight (LoS) path. As a result, the $\alpha-\mu$ fading model degenerates to its limiting case, where the parameter $\mu$ approaches infinity and the amplitude fluctuations vanish. Under this condition, the channel coefficient $g_\mathrm{f}$ can be approximated as a deterministic value, and for normalized channels, it is reasonable to set the fading coefficient $g_\mathrm{f}$ approximately equal to 1, corresponding to $L_\mathrm{f}$ = 0 dB. This simplification is widely adopted in THz link analysis. Therefore, the received power $P_\mathrm{R}$ of the THz signal is calculated to be -57.0 to -53.5 dBm. The link budget in the simulation is demonstrated in Table \ref{table3}.

\begin{table}[!t]
\caption{Link budget\label{table3}}
\centering
\renewcommand{\arraystretch}{1.8}
\begin{tabular}{
>{\centering\arraybackslash}m{1.5cm}
>{\centering\arraybackslash}m{2.5cm}
>{\centering\arraybackslash}m{2.5cm}}
\hline
\textbf{Parameter} & \textbf{Simulation} & \textbf{Experiment} \\
\hline
$P_\mathrm{T}$ & 17.7 $\sim$ 21.2 dBm & 17.7 $\sim$ 21.2 dBm\\
\hline
$G_\mathrm{Tmax}$ & 54 dBi & 54 dBi\\
\hline
$G_\mathrm{Rmax}$ & 57 dBi & 57 dBi\\
\hline
FSPL & -164.1 dB & -164.1 dB\\
\hline
$L_\mathrm{m}$ & -21.6 dB & -21.6 dB\\
\hline
$L_\mathrm{p}$ & 0 dB & -4.7 dB\\
\hline
$L_\mathrm{f}$ & 0 dB & 0 dB\\
\hline
$P_\mathrm{R}$ & -57.0 $\sim$ -53.5 dBm & -61.7 $\sim$ -58.2 dBm\\
\hline
\end{tabular}
\end{table}

At the receiver, the THz signal is collected by a lens and a HA, which together provide a total gain of 57 dBi. The received signal is first amplified by LNA-2 with a gain $G_\mathrm{LNA-2}$ of 38 dB and then down-converted to baseband using an IQ-subharmonic mixer and a LO. The conversion loss $G_\mathrm{Mixer}$ of the mixer is -11 dB. The baseband signal is further amplified by LNA-3, which provides an additional gain $G_\mathrm{LNA-3}$ of 35 dB. The total connection loss of components in the receiver is approximately 1 dB. Finally, the baseband signal is captured by a 100 GSa/s OSC and processed offline using DSP. 

Taking $P_\mathrm{R}$ = -53.5 dBm as an example, the electrical signal power input to the OSC is estimated to be 7.5 dBm (= -53.5+38-11+35-1). The QPSK signal has a peak-to-average power ratio (PAPR) of 0 dB, resulting in a peak power of 7.5 dBm, which corresponds to a $V_\mathrm{pp}$ of 1.05 V. However, in the experiment, the time-domain waveform shown on the OSC has a $V_\mathrm{pp}$ of 0.62 V, which is lower than the estimated 1.05 V. This indicates that, in the above power calculation, the additional power penalties induced by pointing errors should be taken into account. Based on the measured $V_\mathrm{pp}$ of 0.62 V, the average electrical power entering the OSC is calculated to be 2.8 dBm. Therefore, the actual received power $P_\mathrm{R}$ in the experiment is -58.2 dBm (= 2.8-38-35+11+1) and the power loss due to pointing errors is calculated to be 4.7 dB. The link budget in the experiment is demonstrated in Table \ref{table3}.

\subsection{Analysis of pointing errors}
\label{sec:IV.D)}
Over a transmission distance as long as 30 km, achieving precise alignment between the transmitter and receiver antennas becomes impractical. Firstly, the elevation and azimuth angles are calculated from latitude, longitude, altitude, and distance, all of which carry inherent measurement errors. Secondly, the antenna alignment is performed manually during the experiment, which inevitably introduces a small residual pointing offset due to the limited mechanical adjustment precision. These factors collectively lead to additional power loss. Since $L_\mathrm{p}$ originates from the pointing error between antennas, it can be expressed as the attenuation factor associated with the antenna gain, which is expressed as 

\begin{equation}
\label{eq25}
\begin{aligned}
L_\mathrm{p} &= 10 \log_{10} \Bigl[ \exp\bigl( - (\frac{\varphi_\mathrm{a}^\mathrm{T} - \varphi_\mathrm{a}^\mathrm{T'}}{\sigma_\mathrm{a}^\mathrm{T}})^2 
            - (\frac{\varphi_\mathrm{e}^\mathrm{T} - \varphi_\mathrm{e}^\mathrm{T'}}{\sigma_\mathrm{e}^\mathrm{T}})^2 \bigr) \Bigr] \\
&\quad + 10 \log_{10} \Bigl[ \exp\bigl( - (\frac{\varphi_\mathrm{a}^\mathrm{R} - \varphi_\mathrm{a}^\mathrm{R'}}{\sigma_\mathrm{a}^\mathrm{R}})^2 
            - (\frac{\varphi_\mathrm{e}^\mathrm{R} - \varphi_\mathrm{e}^\mathrm{R'}}{\sigma_\mathrm{e}^\mathrm{R}})^2 \bigr) \Bigr].
\end{aligned}
\end{equation}

\begin{table}[!t]
\caption{Antenna angles in the experiment\label{table4}}
\centering
\renewcommand{\arraystretch}{1.8}
\begin{tabular}{
>{\centering\arraybackslash}m{2.5cm}
>{\centering\arraybackslash}m{2cm}}
\hline
\textbf{Parameter} & \textbf{Value} \\
\hline
${\varphi_\mathrm{a}^\mathrm{T}}$ & 0°\\
\hline
${\varphi_\mathrm{e}^\mathrm{T}}$ & -0.27°\\
\hline
${\varphi_\mathrm{a}^\mathrm{R}}$, ${\varphi_\mathrm{a}^\mathrm{R}}’$ & 180°\\
\hline
${\varphi_\mathrm{e}^\mathrm{R}}$, ${\varphi_\mathrm{e}^\mathrm{R}}’$ & 0.27°\\
\hline
${\sigma_\mathrm{a}^\mathrm{T}}$, ${\sigma_\mathrm{e}^\mathrm{T}}$, ${\sigma_\mathrm{a}^\mathrm{R}}$, ${\sigma_\mathrm{e}^\mathrm{R}}$ & 0.3°\\
\hline
${\varphi_\mathrm{a}^\mathrm{T}}’$ & -0.31° $\sim$ 0.31°\\
\hline
${\varphi_\mathrm{e}^\mathrm{T}}’$ & -0.58° $\sim$ 0.04°\\
\hline
\end{tabular}
\end{table}

\begin{figure*}[!t]
\centering
\includegraphics[width=1\linewidth]{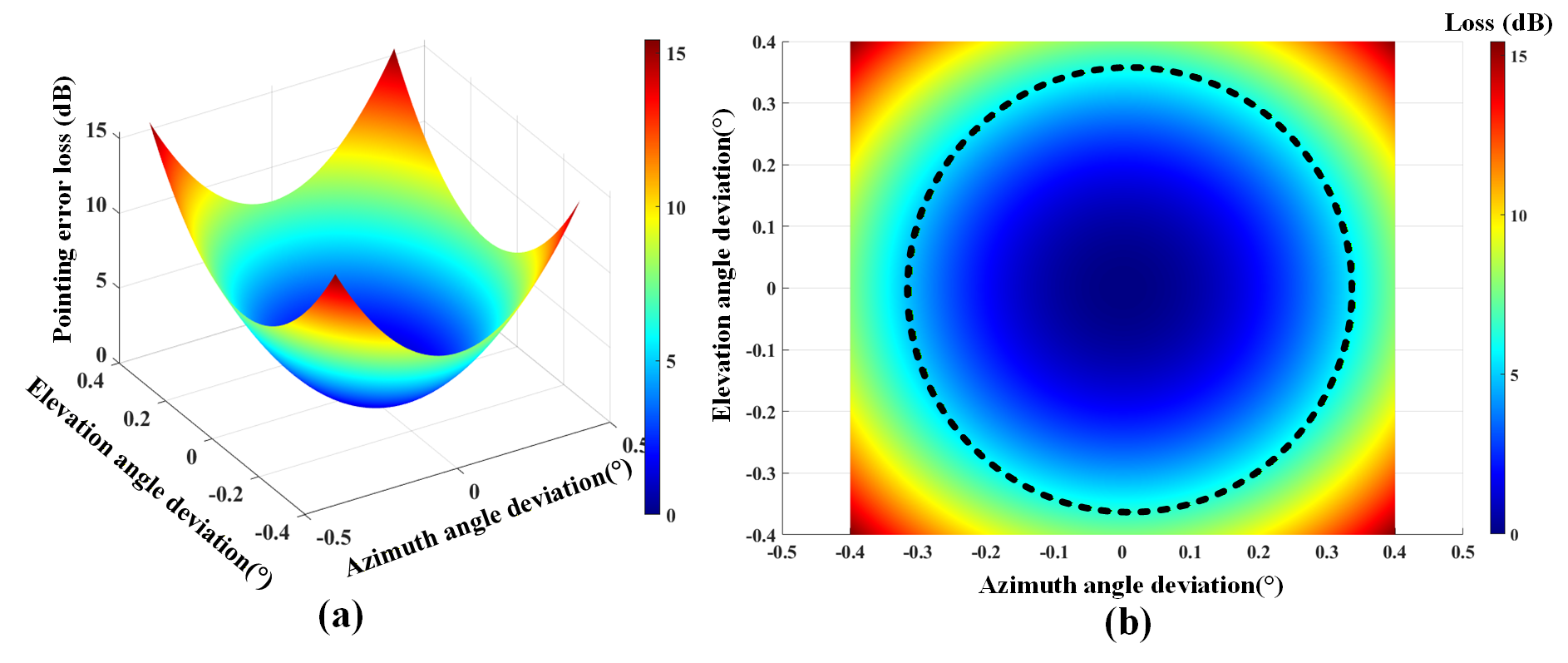}
\caption{(a) The power loss caused by azimuth and elevation angle deviations of the transmitter antenna. (b) The 2D projection of (a).}
\label{fig4}
\end{figure*}

The power loss due to the pointing error is calculated to be 4.7 dB. Using equation \eqref{eq25}, the angle deviation between the transmitter and receiver antennas can be inversely derived. Without loss of generality, it is assumed that the receiver antenna is perfectly aligned, namely ${\varphi_\mathrm{a}^\mathrm{R}}$ = ${\varphi_\mathrm{a}^\mathrm{R}}’$and ${\varphi_\mathrm{e}^\mathrm{R}}$ = ${\varphi_\mathrm{e}^\mathrm{R}}’$. Fig. \ref{fig4}(a) shows the power loss caused by the azimuth and elevation angle deviations of the transmitter antenna. Fig. \ref{fig4}(b) presents the 2D projection of Fig. \ref{fig4}(a). The black dashed lines indicate the ranges of azimuth and elevation deviations that correspond to a power loss of 4.7 dB. Therefore, in the experiment, the deviations of both the azimuth and elevation angles fall within -0.31° to 0.31°, which is attributed to the limited mechanical adjustment precision. This agrees well with the practical experimental conditions, indicating that the THz link model presented in Section \ref{sec:II} shows good consistency with the experimental results. The values of these angles in the experiment are shown in Table \ref{table4}.

\section{Results and Discussion}
\subsection{SNR, BER, and data rate}
\label{sec:V.A)}
\begin{figure*}[!t]
\centering
\includegraphics[width=1\linewidth]{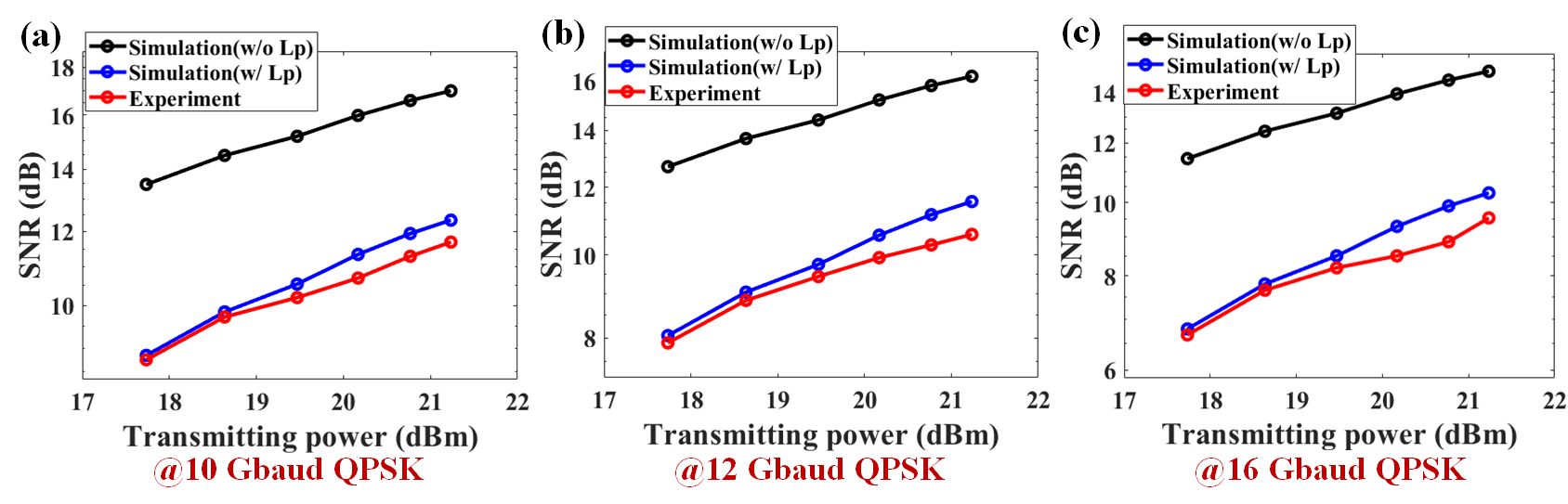}
\caption{Comparison of the SNR between simulation and experiment: (a) 10 Gbaud QPSK, (b) 12 Gbaud QPSK, (c) 16 Gbaud QPSK.}
\label{fig5}
\end{figure*}

Fig. \ref{fig5} presents a comparison of the SNR between simulation and experiment for transmitted QPSK signals with different symbol rates. Since the received power $P_\mathrm{R}$ varies with $P_\mathrm{T}$, Fig. \ref{fig5} presents the relationship between SNR and $P_\mathrm{T}$. Using \eqref{eq9} and the parameters listed in Table \ref{table1}, the equivalent noise figure $NF_\mathrm{R}$ of the THz receiver is calculated to be 3.41 dB. Based on \eqref{eq10} and \eqref{eq24}, the estimated SNR of the received signal without pointing errors is computed, as indicated by the black curve in Fig. \ref{fig5}. This curve represents the SNR predicted by the model and component specifications under ideal conditions and is therefore labeled as “Simulation (w/o pointing errors)" in the legend. As analyzed in Section \ref{sec:IV.C)} and \ref{sec:IV.D)}, the experiment is inherently affected by pointing errors. Such misalignment reduces the received power, thereby decreasing the estimated SNR of the received signal. The SNR in the presence of pointing errors is depicted by the blue curve in Fig. \ref{fig5}. For this curve, the received signal power $P_\mathrm{R}$ is obtained from experimental measurements, while the noise power at the receiver is still determined based on the model and component specifications. Hence, it is labeled as “Simulation (w/ pointing errors)" in the legend. To determine the actual SNR of the received signal, the QPSK signal transmitted in the experiment was captured using an OSC and offline-demodulated to reconstruct the QPSK constellation. The corresponding error vector magnitude (EVM) was then calculated to derive the actual SNR, as indicated by the red curve in Fig. \ref{fig5}. Hence, it is labeled as “Experiment" in the legend.

\begin{figure*}[!t]
\centering
\includegraphics[width=1\linewidth]{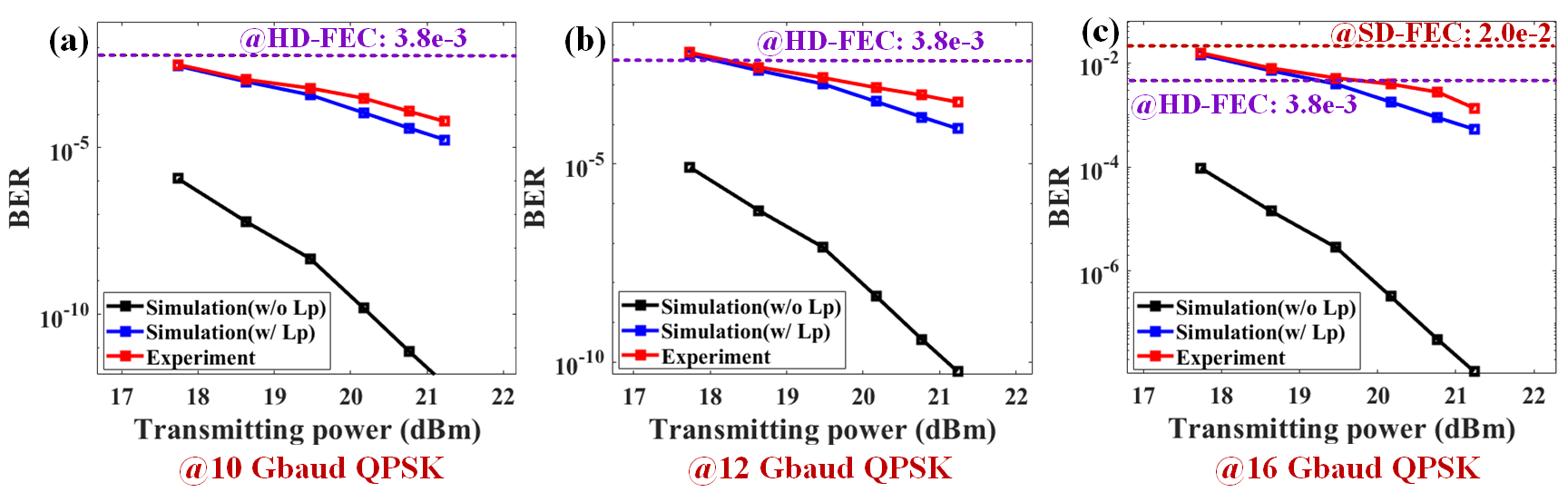}
\caption{Comparison of the BER between simulation and experiment: (a) 10 Gbaud QPSK, (b) 12 Gbaud QPSK, (c) 16 Gbaud QPSK.}
\label{fig6}
\end{figure*}

\begin{figure*}[!t]
\centering
\includegraphics[width=1\linewidth]{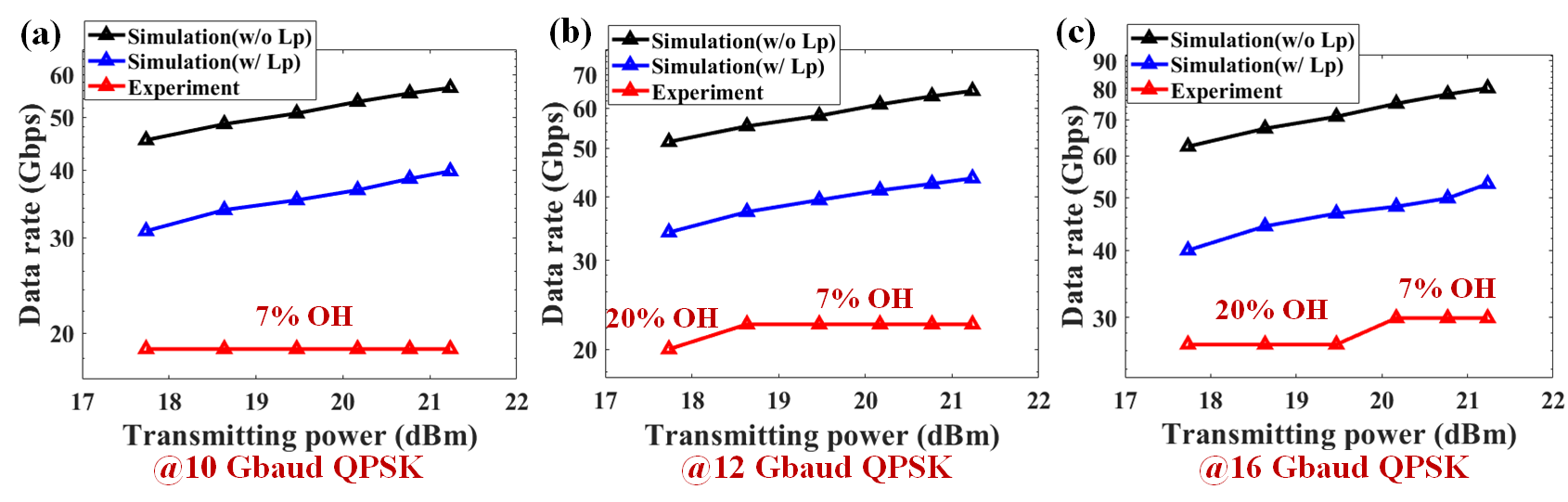}
\caption{Comparison of the data rate between simulation and experiment: (a) 10 Gbaud QPSK, (b) 12 Gbaud QPSK, (c) 16 Gbaud QPSK.}
\label{fig7}
\end{figure*}

Fig. \ref{fig6} presents a comparison of the BER between simulation and experiment for transmitted QPSK signals with different symbol rates. Based on the SNR shown in Fig. \ref{fig5}, the BER in simulation is calculated using the Q-function, as indicated by the black and blue curves in Fig. \ref{fig6}. In contrast, the red curve represents the actual BER measured in the experiment. For the BER represented by the blue curve, both the pointing-error loss and the connection loss of receiver components have been taken into account. Nevertheless, the experimentally measured BER, shown by the red curve, exhibits further degradation. This is because the frequency responses of the devices (e.g., amplifiers and mixer) are not perfectly flat over the 10 GHz signal bandwidth, and additional nonlinear effects are present in the system. As a result, the BER of the offline-demodulated QPSK signal is slightly worse than the value predicted by simulation with pointing errors. Similarly, the experimentally measured SNR is slightly lower than the simulated SNR that includes pointing errors, as shown in Fig. \ref{fig5}.

Fig. \ref{fig7} presents a comparison of the data rate between simulation and experiment for transmitted QPSK signals with different symbol rates. Based on Shannon capacity formula \eqref{Shannon} and SNR in the simulation, the data rate is calculated, as indicated by the black and blue curves in Fig. \ref{fig7}. These two simulation curves indicate the maximum data rate that can be reliably transmitted over a communication channel with a given bandwidth and SNR. It shows the theoretical limit beyond which errors are unavoidable, regardless of coding or modulation. However, in the real experiment, the symbol rate of the signal was fixed, and different SNR and BER values for the QPSK signal were obtained by varying the transmitting power $P_\mathrm{T}$. As a result, the achieved data rate remains constant with increasing $P_\mathrm{T}$, as shown in Fig. \ref{fig7}(a). According to \eqref{rate_SE}, the achievable data rate in the experiment is calculated using spectral efficiency $\eta$, which is determined by the modulation format and the forward error correction (FEC) overhead. For the 10 Gbaud QPSK signal, the pre-FEC BER in Fig. \ref{fig6}(a) remains below the hard-decision forward error correction (HD-FEC) threshold, allowing a 7$\%$ overhead to achieve a post-FEC BER of zero. Therefore, the achievable data rate in the experiment for 10 Gbaud QPSK is 10$\times$2/(1+7$\%$)=18.7 Gbps. Similarly, for the 16 Gbaud QPSK signal, when the transmitting power is below 20 dBm, the BER performance does not meet the HD-FEC threshold and only satisfies the soft-decision FEC (SD-FEC) limit, as shown in Fig. \ref{fig6}(c). In this case, the achievable data rate is 16$\times$2/(1+20$\%$)=26.7 Gbps. As the transmitting power increases, the required FEC overhead decreases, allowing the achievable data rate to increase up to 16$\times$2/(1+7$\%$)=30 Gbps.

\subsection{Power consumption and consumption factor}
\begin{figure}[!t]
\centering
\includegraphics[width=1\linewidth]{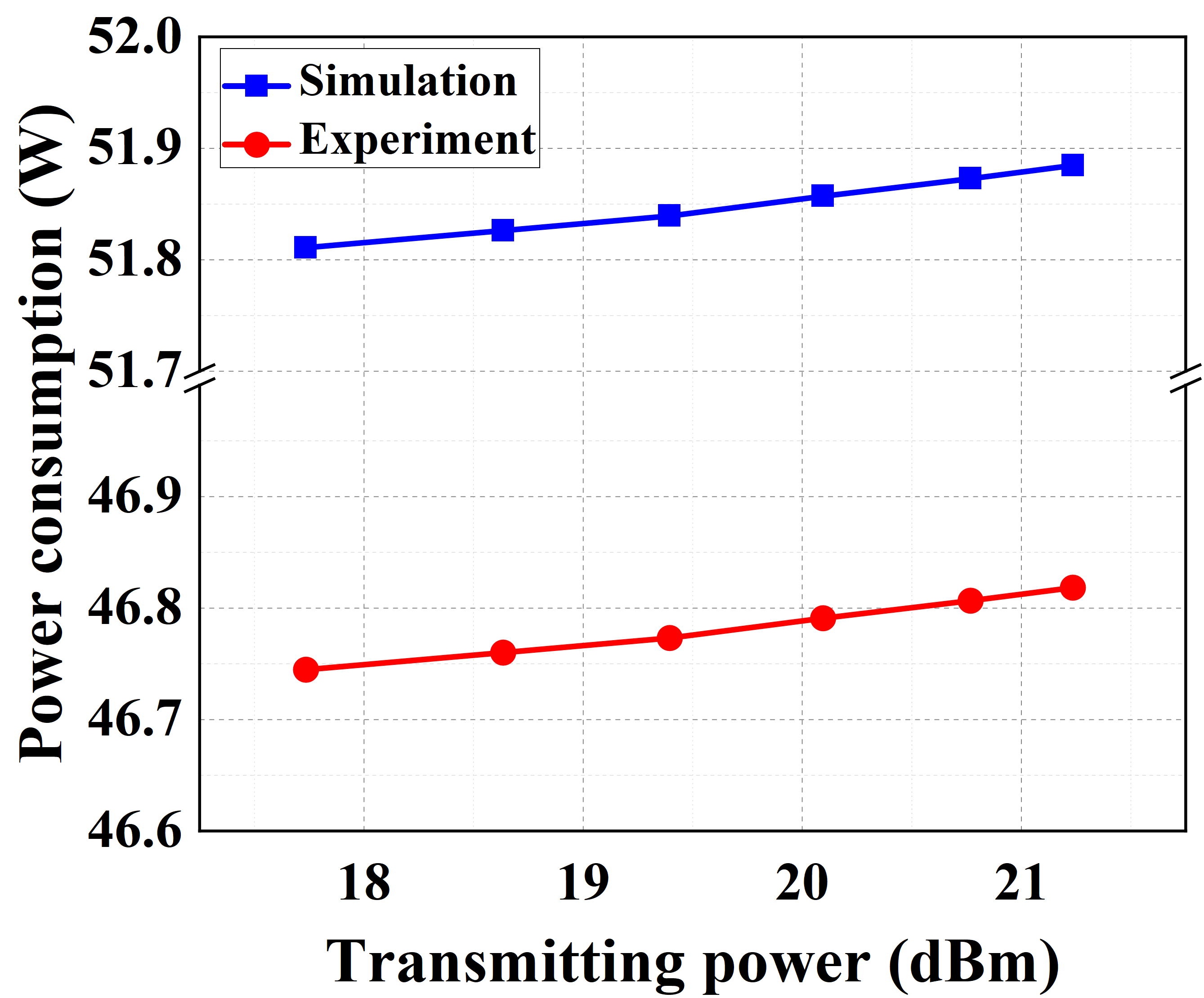}
\caption{Comparison of the total power consumption between simulation and experiment.}
\label{fig8}
\end{figure}

Based on \eqref{eq12}$\sim$\eqref{eq20} and the component datasheets, the power consumption of each component in the simulation is calculated and summarized in Table \ref{table1}. In the experiment, some components, such as amplifiers and the LO, are powered directly by DC supplies. The DC power consumption is calculated based on the voltage and current readings displayed on the power supply output, i.e., $P_\mathrm{DC}$ = $V_\mathrm{sup}$ × $I_\mathrm{sup}$. $V_\mathrm{sup}$ and $I_\mathrm{sup}$ are measured at the output terminals of the DC power supply (Model: Keysight E3631A) using its internal sensing and measurement circuitry and the values correspond to steady-state time-averaged readings under continuous operation. No additional external voltage or current probes are used at the component supply pins. The voltage drop over the wires to the device pins is negligible. However, some components are powered from the mains, and their actual power consumption cannot be directly measured. For these components, the values provided in the datasheets are used instead. A comparison of the total power consumption of the photonic THz transceiver in simulation and experiment is presented in Fig. \ref{fig8}. The measured DC power in the experiment is 46.68 W, which is 5.04 W lower than the DC power consumption of 51.72 W in the simulation. This is because, in the experiment, many components can operate properly even when the actual supply voltage is lower than the value specified in the datasheet. For example, for PA-2, the datasheet specifies a supply voltage of 10 V and a current of 1.45 A. However, in the experiment, the supply voltage of PA-2 is set to 8.6 V. Since PA-2 is bias-regulated, the measured supply current remains unchanged (1.456 A), which reduces the DC power consumption by 10V × 1.45A - 8.6V × 1.456A=1.98 W. The same applies to other components, including the RF driver (0.18 W), IQ modulator (0.656 W), LNA-1/2/3 (0.033/0.01/0.42 W), PA-1 (0.09 W), frequency synthesizer (1.2 W) and frequency multiplier (0.47 W). These are measured values which depend on the internal circuits of the devices, hence out of our control. As a result of these reductions in DC power, the total power consumption measured in the experiment is 5.04 W lower than that in the simulation.

Here, we provide further clarification on the operating condition of PA-2. PA-2 is bias-regulated; therefore, even when the DC supply is reduced to 8.6 V, the transistor still operates in the constant-current region, which keeps the drain current unchanged. Since the saturated output power satisfies $P_\mathrm{sat}$(W) $\propto [V_\mathrm{DD}$(V)]$^2$, where $V_\mathrm{DD}$ is the supply voltage, reducing $V_\mathrm{DD}$ from 10 V to 8.6 V leads to a decrease in $P_\mathrm{sat}$(dBm) of 20$\log_{10}$(10/8.6)=1.3 dB. Therefore, the saturated output power decreases from 23 dBm to 21.7 dBm. In the experiment, even when the input power of PA-2 is set to its maximum value, the maximum output power is 21.2 dBm, which is still below $P_\mathrm{sat}$. The effect of the decrease in $V_\mathrm{DD}$ on the THz output power can be neglected. This is consistent with the observation that the waveform amplitude captured by the OSC at the receiver side remains unchanged. Therefore, the THz output power of PA-2, which ranges from 17.7$\sim$21.2 dBm, is determined solely by the input optical power of the UTC-PD in the experiment. The effect of the reduced supply voltage on the linearity of PA-2 can also be negligible. In terms of power-added efficiency (PAE), because the DC power consumption is reduced while the amplifier gain remains unchanged, the efficiency of PA-2 is improved. Taking an input power of -5.7 dBm as an example, at supply voltages of 10 V and 8.6 V, the PAE of PA-2 is 0.404$\%$ and 0.468$\%$, respectively. The calculation procedure for the PAE is provided in the Appendix

When conducting the experiment, one objective is to minimize the overall system power consumption and extend the components lifetime. Meanwhile, it is observed that reducing the supply voltage resulted in no change in the signal amplitude captured by the OSC, indicating that the SNR of received signals and the BER of demodulated signals were not degraded. Reducing power consumption without compromising signal quality is the primary motivation for lowering the supply voltage.

\begin{figure*}[!t]
\centering
\includegraphics[width=1\linewidth]{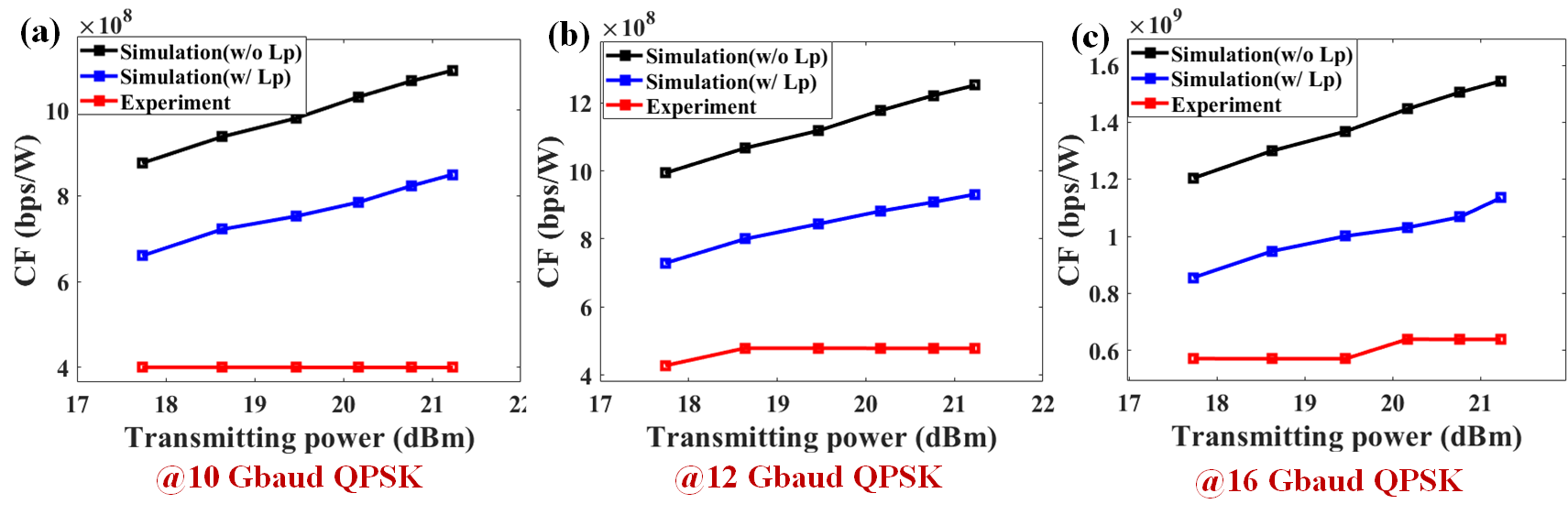}
\caption{Comparison of the consumption factor between simulation and experiment.}
\label{fig9}
\end{figure*}

Based on \eqref{eq11}, \eqref{Shannon} and \eqref{eq12}, the CF in simulation is calculated and expressed in bps/W, as illustrated by the black and blue curves in Fig. \ref{fig9}. Since the simulated data rate in Fig. \ref{fig7} increases much faster than the simulated power consumption shown in Fig. \ref{fig8}, the simulated CF continues to increase with the transmitting power. In the experiment, the actual CF, as shown by the red curve in Fig. \ref{fig9}, is obtained using \eqref{eq11}, \eqref{rate_SE} and experimental power consumption. Since the experimental power consumption in Fig. \ref{fig8} varies only slightly, the variation trend of the experimental CF is mainly determined by the experimental data rate shown in Fig. \ref{fig7}. Additionally, although the power consumption in the experiment is lower than that in the simulation, the data rate is also lower. As a result, the experimental CF performance is slightly worse than the simulated CF, with a reduction of approximately $37\%\sim47\%$.

\subsection{Effect of parameters in the simulation model}
\begin{figure*}[!t]
\centering
\includegraphics[width=1\linewidth]{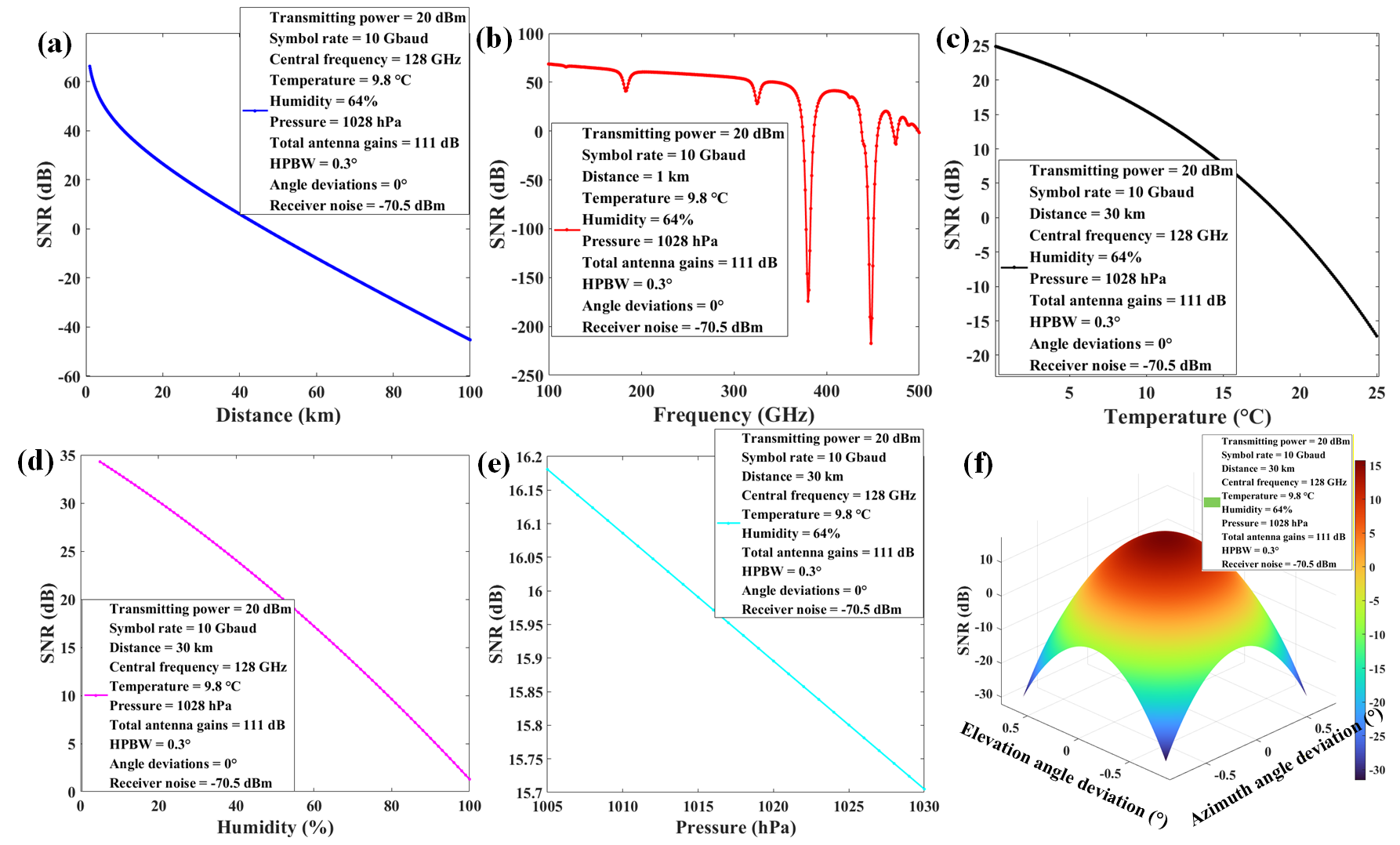}
\caption{The effect of different parameters in the simulation model on the performance of the link: (a) transmission distance, (b) central frequency, (c) temperature, (d) humidity, (e) pressure, and (f) angle deviations.}
\label{fig10}
\end{figure*}
Fig. \ref{fig10} demonstrates the effect of different parameters in the simulation model on the SNR of the received signals. When one parameter is varied, all other parameters are kept constant, and their specific values are listed in the legend of Fig. \ref{fig10}. Since the impacts of transmitting power and symbol rate on system performance have already been analyzed in Section \ref{sec:V.A)}, they are fixed at 20 dBm and 10 Gbaud here. Moreover, antenna gains, HPBW, and receiver noise are inherent characteristics of the hardware components; therefore, their values are also kept constant. It is evident that variations in transmission distance, temperature, humidity, and angle deviation have a more significant impact on link performance. Fig. \ref{fig10} provides a useful simulation-based foundation for link budget analysis and system design in THz communication systems.

\section{Conclusion}
This paper presents the experimental realization of a 30 km 30 Gbps photonic THz wireless communication link. Based on the established 30 km testbed, THz channel models are experimentally assessed for the first time. In addition, the power consumption of the photonic THz communication system is analyzed, with a detailed comparison between simulation results and experimental measurements. Furthermore, system performance is assessed in terms of SNR, BER, data rate and CF, allowing a direct examination of the agreement between simulation and experimental results. Future work will extend the present study by investigating the performance of THz communication system under different weather conditions and over different link distances. Beyond standalone THz links, the system will be integrated with free-space optical (FSO) communication to realize a hybrid THz/FSO architecture, aiming to leverage the complementary advantages of both technologies in terms of reliability, capacity, and environmental adaptability.

\section*{Acknowledgments}
The authors would like to thank Prof. Harald Haas, Hossein Safi, Othman Younus, Mingqing Liu and Tongyun Li for their valuable discussions, constructive suggestions, and technical support during the course of this work while the first author was visiting the University of Cambridge in the UK. 

\appendix[Comparison between two THz link models]
Besides the THz link model\eqref{eq1}$\sim$\eqref{eq7} introduced in Section \ref{sec:II}, Ref.\cite{Ref.63} also presents another THz link model. The path loss, as given in \eqref{eq26}, includes the gains of both the transmitting and receiving antennas,
\begin{equation}
\label{eq26}
\begin{alignedat}{1}
g_\mathrm{a} =\frac{c\sqrt{ G_\mathrm{Tmax}G_\mathrm{Rmax} }}{4 \pi f d} 
       \, \exp\Biggl[- \frac{1}{2} \kappa_\alpha(f) \, d \Biggr].
\end{alignedat}
\end{equation}
The misalignment effect adopts the same pointing error model as that used in FSO links \cite{Ref.64}, as shown in \eqref{eq27}, where the alignment degradation is quantified by the reduction in received power caused by the offset between the THz Gaussian beam footprint and the receiver center.
\begin{equation}
\label{eq27} 
g_\mathrm{m} \approx A_{0} \, 
        \exp\Biggl(- \frac{2 \, r_\mathrm{d}^2}{\omega_\mathrm{eq}^2} \Biggr), 
        \quad \text{for } r_\mathrm{d} \ge 0,
\end{equation}
where $A_{0}$ represents the fraction of power collected at the detector when there are no pointing errors, $r_\mathrm{d}$ represents the radial displacement between the beam center and the detector, and $\omega_\mathrm{eq}$ denotes the equivalent beamwidth. $A_{0}$ and $\omega_\mathrm{eq}$ are defined as

\begin{equation}
\label{eq28}
\left\{
\begin{aligned}
A_{0} &= \bigl[ \operatorname{erf}(v) \bigr]^2, \\[4pt]
\omega_\mathrm{eq}^2 &= \frac{\omega_\mathrm{z}^2 \sqrt{\pi} \, \operatorname{erf}(v)}{2 v \, \exp(-v^2)}, \\[4pt]
v &= \sqrt{\frac{\pi}{2}} \, \frac{r_\mathrm{a}}{\omega_\mathrm{z}},
\end{aligned}
\right.
\end{equation}
where erf(·) represents the error function, $r_\mathrm{a}$ denotes the receiver aperture and $\omega_\mathrm{z}$ is the Gaussian beamwaist at the receiver, which is defined as
\begin{equation}
\label{eq29}
\omega_\mathrm{z} = \omega_0 
\sqrt{
  1 + 
  \left[
    1 + \frac{2 \, \omega_0^2}{
      \bigl( 0.55 \, C_\mathrm{n}^2 \, k_0^2 \, d \bigr)^{-6/5}
    }
  \right]
  \left( \frac{\lambda \, d}{\pi \, \omega_0^2} \right)^2
},
\end{equation}
where $\omega_{0}$ the primary beam waist radius. Note that the multipath fading of the THz link is also modeled using an $\alpha-\mu$ distribution, the same as in Section \ref{sec:II.C}.

Based on the model\eqref{eq26}$\sim$\eqref{eq29}, the received power $P_\mathrm{R}$ can be estimated and compared with the model\eqref{eq1}$\sim$\eqref{eq7} and experimental results, as shown in Table \ref{table5}. Obviously, the received power predicted by is far below the receiver noise floor, which deviates significantly from the experimental results.
\begin{table}[!t]
\caption{Comparison between two models and experiment\label{table5}}
\centering
\renewcommand{\arraystretch}{1.8}
\begin{tabular}{
>{\centering\arraybackslash}m{1.5cm}
>{\centering\arraybackslash}m{1.8cm}
>{\centering\arraybackslash}m{1.8cm}
>{\centering\arraybackslash}m{1.5cm}}
\hline
\textbf{Parameter} & \textbf{Model\eqref{eq1}$\sim$\eqref{eq7}} & \textbf{Model\eqref{eq26}$\sim$\eqref{eq29}} & \textbf{Experiment} \\
\hline
$P_\mathrm{T}$ (dBm) & 21.2 & 21.2 & 21.2\\
\hline
$g_\mathrm{a\_f}$ (dB) & -164.1 & -53.1 & N/A \\
\hline
$g_\mathrm{a\_m}$ (dB) & -21.6 & -21.6 & N/A \\
\hline
$g_\mathrm{m}$ (dB) & 111 & -52.2 & N/A \\
\hline
$g_\mathrm{f}$ (dB) & 0 & 0 & N/A \\
\hline
$P_\mathrm{R}$ (dBm) & -53.5 & -105.7 & -58.2\\
\hline
\end{tabular}
\end{table}
In fact, in \eqref{eq27}, the misalignment effect accounts for the power loss arising from both geometric spreading and pointing errors, while the path loss in \eqref{eq26} is also derived based on geometric propagation. As a result, the power loss caused by beam divergence is effectively counted twice. This can be further verified by the fact that the values of $g_\mathrm{a\_f}$ and $g_\mathrm{m}$ in model\eqref{eq26}$\sim$\eqref{eq29} are very close, as shown in Table \ref{table5}. This double-counting of beam-divergence loss explains why the results predicted by model\eqref{eq26}$\sim$\eqref{eq29} deviate significantly from the experimental observations. In fact, some papers \cite{Ref.63,Ref.65,Ref.66,Ref.67,Ref.68} ignored the double-counting of beam-divergence loss and used the model\eqref{eq26}$\sim$\eqref{eq29} wrongly. These papers didn't validate the THz channel model using a fully implemented THz communication experiment.

\appendix[Calculation procedure for the PAE]
The PAE is defined as
\begin{equation}
\label{eq30}
{\rm{PAE}}= \frac{{{P_{{\rm{out}}}} - {P_{{\rm{in}}}}}}{{{P_{{\rm{DC}}}}}} \times 100\% ,
\end{equation}
where $P_\mathrm{out}$ represents the output power, $P_\mathrm{in}$ denotes the input power and $P_\mathrm{DC}$ denotes the DC power consumption, all measured in Watts. Taking an input power of -5.7 dBm and an output power of 17.7 dBm as an example, the PAE calculation procedure at supply voltages of 10 V and 8.6 V is given as follows,

\begin{equation}
\label{eq31}
\left\{
\begin{aligned}
\mathrm{PAE}_{10\mathrm{V}} 
&= \frac{10^{(17.7-30)/10} - 10^{(-5.7-30)/10}}
{10 \times 1.45} \times 100\% \\
&= 0.404\% ,\\[1ex]
\mathrm{PAE}_{8.6\mathrm{V}} 
&= \frac{10^{(17.7-30)/10} - 10^{(-5.7-30)/10}}
{8.6 \times 1.456} \times 100\% \\
&= 0.468\% .
\end{aligned}
\right.
\end{equation}

\bibliographystyle{IEEEtran}
\bibliography{reference}

\end{document}